\documentclass{aa}
\usepackage{graphicx,lscape}
\usepackage{deluxetable}
\usepackage{pdflscape}
\usepackage{ulem}
\usepackage{color}

\usepackage{txfonts}
\usepackage{natbib}
\usepackage{longtable}
\bibpunct{(}{)}{;}{a}{}{,}

\definecolor{red}{rgb}{0.75,0.0,0.0}
\definecolor{yel}{rgb}{0.65,0.65,0.0}
\definecolor{grn}{rgb}{0.0,0.75,0.0}
\definecolor{blu}{rgb}{0.0,0.0,0.75}
\definecolor{gry}{rgb}{0.75,0.75,0.75}

\begin{document}

\def\hi{H\,{\sc i}}
\def\hii{H\,{\sc ii}}
\def\hei{He\,{\sc i}}
\def\heii{He\,{\sc ii}}
\def\cii{C\,{\sc ii}}
\def\ciii{C\,{\sc iii}}
\def\civ{C\,{\sc iv}}
\def\niii{N\,{\sc iii}}
\def\nv{N\,{\sc v}}
\def\oi{O\,{\sc i}}
\def\oiii{O\,{\sc iii}}
\def\oiv{O\,{\sc iv}}
\def\ov{O\,{\sc v}}
\def\ovi{O\,{\sc vi}}
\def\ovii{O\,{\sc vii}}
\def\mgii{Mg\,{\sc ii}}
\def\siiv{Si\,{\sc iv}}
\def\siii{Si\,{\sc ii}}
\def\siiii{Si\,{\sc iii}}
\def\siv{S\,{\sc iv}}
\def\svi{S\,{\sc vi}}
\def\alii{Al\,{\sc ii}}
\def\aliii{Al\,{\sc iii}}
\def\piii{P\,{\sc iii}}
\def\pv{P\,{\sc v}}
\def\feii{Fe\,{\sc ii}}
\def\feiii{Fe\,{\sc iii}}
\def\neviii{Ne\,{\sc viii}}
\def\mgx{Mg\,{\sc x}}
\def\nh{\ifmmode n_\mathrm{\scriptscriptstyle H} \else $n_\mathrm{\scriptscriptstyle H}$\fi}
\def\ne{\ifmmode n_\mathrm{\scriptscriptstyle e} \else $n_\mathrm{\scriptscriptstyle e}$\fi}
\def\qh{\ifmmode Q_\mathrm{\scriptscriptstyle H} \else $Q_\mathrm{\scriptscriptstyle H}$\fi}
\def\uh{\ifmmode U_\mathrm{\scriptscriptstyle H} \else $U_\mathrm{\scriptscriptstyle H}$\fi}
\def\Nh{\ifmmode N_\mathrm{\scriptscriptstyle H} \else $N_\mathrm{\scriptscriptstyle H}$\fi}
\def\mdot{$\dot{M}$}
\def\edot{$\dot{E}_\mathrm{k}$}
\def\lbol{$L_\mathrm{Bol}$}
\def\ledd{$L_\mathrm{Edd}$}
\def\Zsun{\ifmmode {\rm Z}_{\odot} \else Z$_{\odot}$\fi}
\def\Msun{\ifmmode {\rm M}_{\odot} \else M$_{\odot}$\fi}
\def\kms{\ifmmode {\rm km~s}^{-1} \else km~s$^{-1}$\fi}
\def\Lya{\ifmmode {\rm Ly}\alpha \else Ly$\alpha$\fi}
\def\Lyb{\ifmmode {\rm Ly}\beta \else Ly$\beta$\fi}
\def\Lyg{\ifmmode {\rm Ly}\gamma \else Ly$\gamma$\fi}
\def\Lyd{\ifmmode {\rm Ly}\delta \else Ly$\delta$\fi}
\newcommand{\secref}[1]{Section~\ref{#1}}
\newcommand{\chpref}[1]{Chapter~\ref{#1}}
\newcommand{\figref}[1]{Figure~\ref{#1}}
\newcommand{\tabref}[1]{Table~\ref{#1}}
\newcommand{\eqnref}[1]{Equation~(\ref{#1})}

\def\gtorder{\mathrel{\raise.3ex\hbox{$>$}\mkern-14mu
             \lower0.6ex\hbox{$\sim$}}}
\def\ltorder{\mathrel{\raise.3ex\hbox{$<$}\mkern-14mu
             \lower0.6ex\hbox{$\sim$}}}
\def\proptwid{\mathrel{\raise.3ex\hbox{$\propto$}\mkern-14mu
             \lower0.6ex\hbox{$\sim$}}}


\title{Multi-wavelength campaign on NGC 7469 \\ V. Analysis of the HST/COS observations: \\ Super solar metallicity, distance, and trough variation models}

\author{N. Arav\inst{1}
  \and
	X. Xu\inst{1}
	\and
	G.A. Kriss\inst{2}
	\and
	C. Chamberlain\inst{1}	
	\and
	T. Miller\inst{1}
	\and
	E. Behar\inst{3}
	\and
	J.S. Kaastra\inst{4,5}
	\and
	J.C. Ely\inst{2}
	\and
	U. Peretz\inst{3}
	\and
	M. Mehdipour\inst{4}
	\and
	G. Branduardi-Raymont\inst{6}
	\and
	S. Bianchi\inst{7}
	\and
	M. Cappi\inst{8}
	\and
	E. Costantini\inst{4}
	\and
	B. De Marco\inst{9}
	\and
	L. di Gesu\inst{10}
	\and
	J. Ebrero\inst{11}
	\and
	S. Kaspi\inst{3,12}
	\and
	R. Middei\inst{7}
	\and
	P.-O. Petrucci\inst{13}
	\and
	G. Ponti\inst{14,15}
	}

\offprints{arav@vt.edu}

\institute{
	Department of Physics, Virginia Tech, Blacksburg, VA 24061, USA.
	\and
	Space Telescope Science Institute, 3700 San Martin Drive, Baltimore, MD 21218, USA.
	\and
	Department of Physics, Technion-Israel Institute of Technology, 32000 Haifa, Israel.
	\and
	SRON Netherlands Institute for Space Research, Sorbonnelaan 2, 3584 CA Utrecht, the Netherlands.
	\and
	Leiden Observatory, Leiden University, Post Office Box 9513, 2300 RA Leiden, Netherlands.
	\and
	Mullard Space Science Laboratory, University College London, Holmbury St. Mary, Dorking, Surrey, RH5 6NT, UK.
	\and
	Dipartimento di Matematica e Fisica, Universit\`{a} degli Studi Roma Tre, via della Vasca Navale 84, 00146 Roma, Italy.
	\and
	INAF-IASF Bologna, Via Gobetti 101, I-40129 Bologna, Italy.
	\and
	N. Copernicus Astronomical Center of the Polish Academy of Sciences,
Bartycka 18, 00-716 Warsaw
	\and
	Italian Space Agency (ASI), Via del Politecnico snc, Rome, Italy 
	\and
	European Space Astronomy Centre, PO Box 78, 28691 Villanueva de la Cañada, Madrid, Spain
	\and
	School of Physics and Astronomy and Wise Observatory, Raymond and Beverly
Sackler Faculty of Exact Sciences, Tel Aviv University, Tel Aviv 6997801, Israel
	\and
	Univ. Grenoble Alpes, CNRS, IPAG, 38000 Grenoble, France
	\and
INAF, Osservatorio Astronomico di Brera Merate, via E. Bianchi 46, I-23807 Merate, Italy
\and
Max-Planck-Institut f\"{u}r extraterrestrische Physik, Giessenbachstrasse, D-85748 Garching, Germany.
}

\date{\today}

\abstract
{{AGN outflows are thought to influence the evolution of their host galaxies and their super massive black holes. To better understand these outflows, we executed a  deep multiwavelength campaign on NGC 7469.
The resulting data, combined with those of earlier epochs, allowed us to construct a comprehensive physical, spatial, and temporal picture for this AGN wind.}}
{{Our aim is to determine the  distance of the UV outflow components from the central source, their abundances and total column-density, and the mechanism responsible for their observed absorption variability.}} 
{We studied the UV spectra acquired during the campaign as well as from  three previous epochs (2002-2010). Our main analysis tools are ionic column-density extraction techniques and  photoionization models (both equilibrium and time-dependent models) based on the code Cloudy.} 
{{For component 1 (at --600 \kms) our findings include the following: metallicity that is roughly twice solar; a simple model based on a fixed total column-density absorber, reacting to changes in ionizing illumination that matches the  different ionic column densities derived from four spectroscopic epochs spanning 13 years; and a distance of   R=$6^{+2.5}_{-1.5}$ pc from the central source.  Component 2 (at --1430~\kms) has shallow troughs and is at a much larger $R$.
For component 3 (at --1880 \kms) our findings include: a similar metallicity to component 1; a photoionization-based model can explain the major features of its complicated absorption trough variability and an upper limit of 60 or 150 pc on $R$. This upper limit is consistent and complementary to the X-ray derived lower limit of 12 or 31 pc for $R$.  The total column density of the UV phase is roughly 1\% and 0.1\% of  the lower and upper ionization components of the warm absorber, respectively.}}
{{The NGC 7469 outflow shows super-solar metallicity similar to the outflow in Mrk 279, carbon and nitrogen are twice and four times more abundant than their solar values, respectively.  Similar to the NGC 5548 case, a simple model can explain the physical characteristics and the variability observed in the outflow.}}


\titlerunning{NGC 7469}
\authorrunning{N. Arav et al.}

\maketitle

\section{Introduction}\label{secIntro}

Absorption outflows are detected as blueshifted troughs in the rest-frame spectrum of active galactic nuclei (AGN). Such outflows in powerful quasars can expel sufficient gas from their host galaxies to halt star formation, limit their growth, and lead to the co-evolution of the size of the host and the mass of its central super massive black hole \citep[e.g.,][]{Ostriker10,Hopkins10,Soker11,Ciotti10,Faucher-Giguere12,Borguet13,Arav13}. Therefore, deciphering the properties of AGN outflows is crucial for testing their role in galaxy evolution. 

Nearby bright Seyfert I objects are important laboratories for studying these outflows because they yield the following: high-resolution
 high-resolution UV data, which allow us to study the outflow kinematics, trough variability, and can yield diagnostics for their distance from the central source; and X-ray grating spectra that give the physical conditions for the bulk of the outflowing material 
\citep[e.g.,][]{Steenbrugge05,Gabel05b,Arav07,Costantini07,Kaastra12,Kaastra14, Arav15}. Therefore, such observations are a crucial stepping stone for quantifying outflows from the luminous (but distant) quasars, for which grating X-ray data are seldom available.

NGC 7469 exhibits three kinematically-distinct UV outflow absorption components with velocity centroids at $-540~\kms$, $-1430~\kms$, and $-1880~\kms$ for components 1, 2 and 3, respectively.
The outflow in NGC 7469 has been previously studied by \cite{Kriss03} and \cite{Scott05}, using UV data from FUSE and STIS as well as simultaneous X-ray observations from Chandra and XMM-Newton.
The UV analysis from \cite{Kriss03} examined a single epoch of FUSE spectra from 1999.
\cite{Scott05} obtained followup FUSE observations of NGC 7469 in 2002, in addition to STIS spectra in 2002 and again in 2004.
These multi-epoch spectra revealed trough variability in all of the detected troughs in components 1 and 3 of the outflow.

To further explore the variability of the NGC 7469 outflow and
establish its location and physical characteristics, we carried out a multiwavelength campaign in 2015 
similar to our previous successful campaigns on
Mrk 509 \citep{Kaastra11a, Mehdipour11, Kriss11b, Arav12, Kaastra12} and
NGC 5548 \citep{Kaastra14, Arav15}.
The 2015 campaign used {\it XMM-Newton} \citep{Behar17},
{\it NuSTAR} \citep{Middei18},
{\it Chandra} and {\it Swift} \citep{Mehdipour18},
and the {\it Hubble Space Telescope} (HST) (this paper) to monitor
NGC 7469 over a six-month interval.
Our seven visits with {\it XMM-Newton} \citep{Behar17} revealed a
photoionized X-ray wind at outflow velocities of
--550, --950,~and~ --2050 \kms\ that had not changed significantly in
its properties since the prior
{\it XMM-Newton} observations in 2000 \citep{Blustin03}
and 2004 \citep{Blustin07},
and the {\it Chandra} observation in 2002 \citep{Scott05}.
The new, deeper {\it XMM-Newton} spectra also revealed emission from
photoionized gas at $-450~\rm km~s^{-1}$, which is compatible with being the
same outflow producing the absorption.
The lack of X-ray absorption-line variability on timescales of roughly a decade
place the X-ray outflow at distances of $>$12--31 pc \citep{Peretz18}. This is
possibly compatible with the 1-kpc starburst ring surrounding the nucleus
\citep{Wilson91}, which is also detected in our X-ray observations
\citep{Mehdipour18}.

This paper analyzes the HST component of our campaign in detail, while using inferences from prior UV observations and all X-ray epochs.
In \secref{secObsDesc} we describe the observations and define the epochs of spectral observation considered in the analysis.
In \secref{secAnalysis} we describe   the unabsorbed emission model and extract column density measurements from the absorption features.
In \secref{secPhoto}  we model the photoionization structure of the outflow components and
their variability.
In section 5 we compare the physical parameter of the UV outflow with those inferred for the warm absorber. 
We summarize our findings, and compare them with other Seyfert outflows in \secref{secSummary}.
\section{Description of Observations}\label{secObsDesc}

Previous to our 2015 campaign, NGC 7469 (J2000: RA=23 03 15.62, DEC=+08 52 26.4, z=0.016268) was observed twice by FUSE (1999 and 2002) and with the
Hubble Space Telescope (HST) in three spectral epochs: 2002 (PID 9095), 2004 (PID 9802) and 2010 (PID 12212). Details of the UV spectral observations are given in \tabref{tabObs}.

The main component of our 2015 campaign to monitor the outflowing absorbing gas in NGC 7469
used seven observations with the Cosmic Origins Spectrograph (COS) on HST (PID 14054) to sample timescales
ranging from 1 day to 6 months.
Each of our seven observations were coordinated with simultaneous
{\it XMM-Newton} observations as described in \citet{Behar17}.
These COS spectra consist of two-orbit visits using gratings G130M and G160M
to cover the 1130--1800 \AA\ wavelength range at a nominal
resolving power of $\sim$18,000 \citep{Green12}.
We also obtained two additional observations (PID 14242) using the blue-mode
of grating G130M at a central wavelength setting of 1096 \AA.
These observations span wavelengths 930--1280 \AA\ (with a gap from
1080--1100 \AA) at a resolving power of $\sim$12,000 \citep{Debes16}, and cover the \ovi\ and \Lyb\ outflow troughs.

The first visit in 2015 occurred five months before the remaining visits, and we distinguish this visit as the 2015a epoch, whereas visits 2-9 were coadded and hereafter referred to as the 2015b epoch. The 2015b epoch spans $\Delta t=33.4$ days of observations which begin $\Delta t=164.8$ days after the 2015a epoch.

\begin{table}
\caption{Observation information for all epochs}\label{tabObs}
\begin{tabular}{llll@{\hspace{-0.9cm}}r}
\multicolumn{1}{c}{Epoch}	&\multicolumn{1}{c}{Obs Date}	&\multicolumn{1}{c}{Instrument}	&\multicolumn{1}{c}{Grating}	&\multicolumn{1}{c}{Exp}\\\hline
1999		&1999/12/06	&FUSE		&FUV	&37.6ks\\
2002		&2002/12/13	&FUSE		&FUV	&7.0ks\\
		&2002/12/13	&HST:STIS	&E140M	&13.0ks\\
2004		&2004/06/21	&HST:STIS	&E140M	&22.8ks\\
2010		&2010/10/16	&HST:COS	&G130M	&2.1ks\\
		&		&		&G160M	&2.4ks\\
2015\_v1$^a$	&2015/06/12	&HST:COS	&G130M	&2.2ks\\
		&		&		&G160M	&2.4ks\\
2015\_v2$^b$	&2015/11/24	&HST:COS	&G130M	&2.2ks\\
		&		&		&G160M	&2.4ks\\
2015\_v3$^b$	&2015/12/15	&HST:COS	&G130M	&2.2ks\\
		&		&		&G160M	&2.4ks\\
2015\_v4$^b$	&2015/12/22	&HST:COS	&G130M$^c$	&12.9ks\\
2015\_v5$^b$	&2015/12/23	&HST:COS	&G130M	&2.2ks\\
		&		&		&G160M	&2.4ks\\
2015\_v6$^b$	&2015/12/25	&HST:COS	&G130M	&2.2ks\\
		&		&		&G160M	&2.4ks\\
2015\_v7$^b$	&2015/12/26	&HST:COS	&G130M	&2.2ks\\
		&		&		&G160M	&2.2ks\\
2015\_v8$^b$	&2015/12/27	&HST:COS	&G130M$^c$	&10.2ks\\
2015\_v9$^b$	&2015/12/29	&HST:COS	&G130M	&2.2ks\\
		&		&		&G160M	&2.4ks\\\hline
\multicolumn{5}{l}{\parbox[t]{0.9\columnwidth}{$^a$Used as 2015a epoch. $^b$Coadded as 2015b epoch.}}\\
\multicolumn{5}{l}{\parbox[t]{0.9\columnwidth}{$^c$Using the 1096 central wavelength setting.}}\\
\end{tabular}
\end{table}

\begin{figure*}
\includegraphics[height=\textwidth,angle=90]{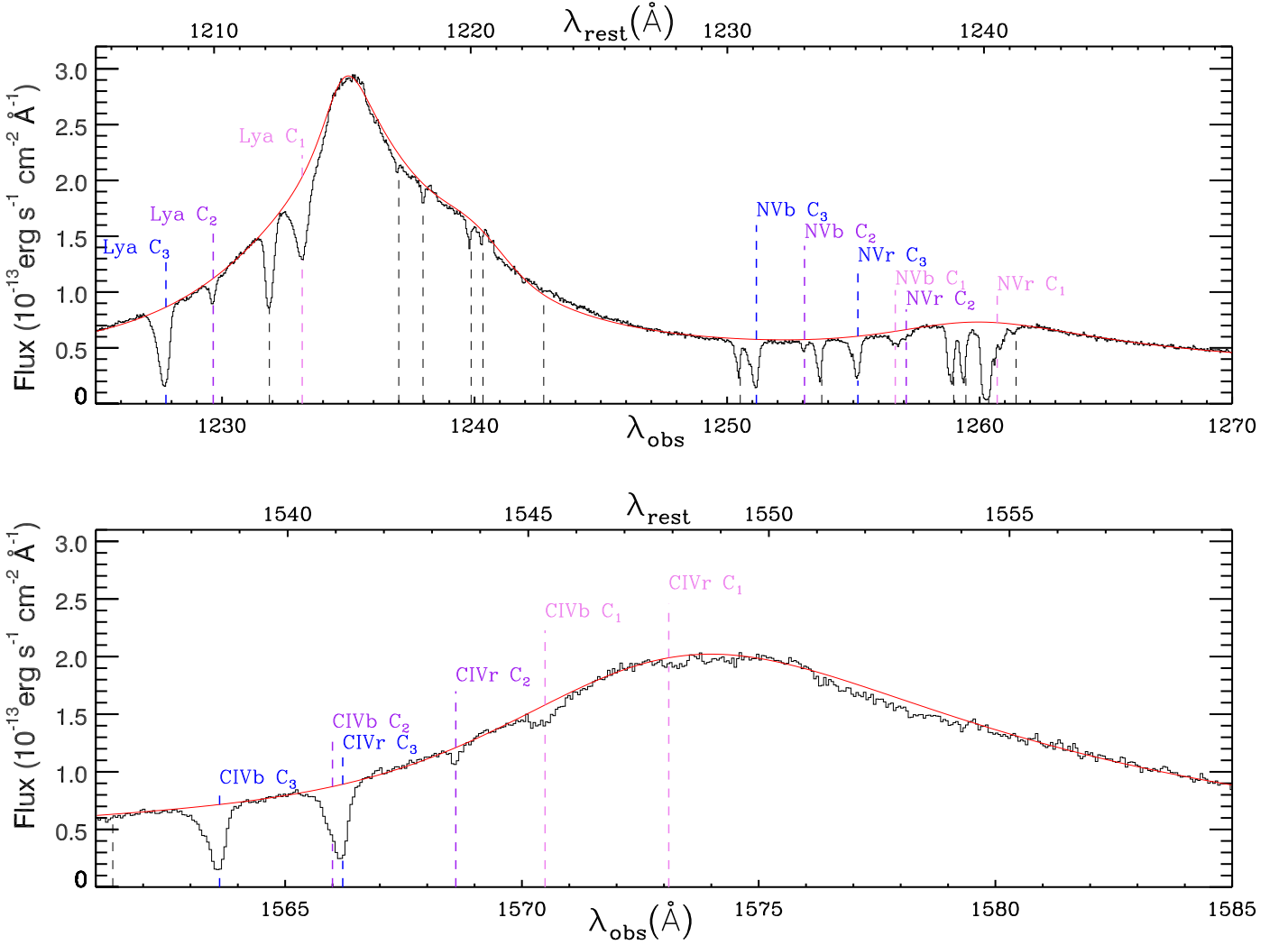}
\caption{HST/COS spectrum of NGC 7469 for the 2015b epoch covering the main outflow absorption features. We label the ionic absorption troughs for all 3 outflow components (C$_1$--C$_3$) from \Lya, \civ\ and \nv;
NVb and NVr refer to the blue and red doublet components of the \nv\ doublet, respectively, With similar meaning for CIVb and CIVr.  Intervening absorption troughs are marked with black dashed lines.
The continuum and BEL emission model is shown as a red solid line.}
\label{figSpecPlot}
\end{figure*}

\section{Spectral Analysis}\label{secAnalysis}
\begin{figure}
\includegraphics[height=\columnwidth,angle=-90]{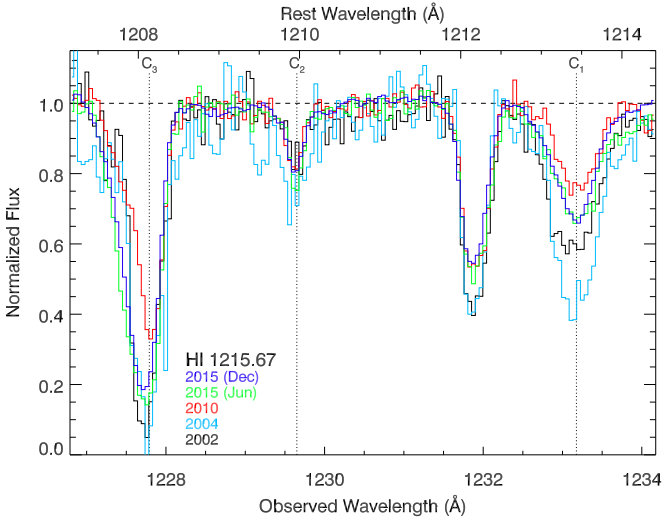}
\caption{Normalized spectrum of the \Lya\ absorption for the three outflow components during the five major observation epochs. The trough at 1232 \AA\ is a \Lya\ intervening absorber.}
\label{figEpochNormLyaAllComp}
\end{figure}

 \figref{figSpecPlot} shows data from the coadded 2015b epoch (see Table 1). The spectral regions in the plot cover the three absorption components as seen in \Lya, \nv\ and \civ.
Components 3 and 1 have a substantial optical depth and exhibit significant variability in multiple ions between multiple epochs (see \figref{figEpochNormLyaAllComp}). Therefore, we focus our study on those components. Component 2 has shallow troughs which make the analysis results more uncertain (see section 4.2). 
The absorption feature at the  1232~\AA\ observed wavelength (see \figref{figEpochNormLyaAllComp})  has a narrow profile and no
metal-line absorption counterparts. In addition, the  1232~\AA\ trough has the same depth in the 2 STIS epochs and the same depth (although shallower than the STIS depth) in the 3 COS epochs. We attribute the depth difference between the STIS and COS epochs to the different resolution and line-spread-function of the two instruments and not to real optical depth change.  Therefore, we agree with 
\cite{Kriss03} and \cite{Scott05}
who attributed the  1232~\AA\ trough to intergalactic Ly$\alpha$
absorption. 

Two of the outflow components (1 and 3) were previously seen in HST spectra of NGC 7469 by
\cite{Kriss03} and \cite{Scott05}. Component 2 was
too weak to be noticeable in the lower S/N STIS spectra from 2002 and 2004.
 Component 2 is clearly visible in our new COS spectra (as well
as in the archival 2010 spectrum), and  it  varies. We conclude that component 2 is part of the 
 NGC 7469 outflow. 
\figref{figEpochNormLyaAllComp} shows the \Lya\ trough variability for all three components.

\subsection{Emission Model}\label{secEmit}
We model the emission separately for each epoch.
The continuum is fit using a single power law, and the broad and narrow line emission features are fit ad-hoc with a superposition of one to four Gaussians per feature.
The total emission model is shown in \figref{figSpecPlot}.
The emission in the spectral regions near the absorption troughs of component 3 is well-defined, but the lower velocity of component 1 locates its absorption troughs close to the corresponding peaks of the emission features.
As such, the absorption depth of component 1 is affected by the emission model, and we assign to its measurements a larger uncertainty as a result.

\subsection{Column Density Extraction}\label{secColDen}
We extract the ionic column densities from each component for every epoch using the partial-covering (PC) method described in \cite{Arav08} for the doublet transitions of \civ\ and \nv.
This method is performed assuming separate covering in each velocity bin.
The remaining ions either have only one absorption trough (\hi), are too shallow (\siiv), or too saturated (\ovi) to use the PC method; so we measure the column densities using the apparent optical depth (AOD) method instead.
We treat the AOD measurements of \hi\ and \ovi\ (when available) as lower limits.
\siiv\ does not appear in the majority of epochs, so we determine upper limits to the column density in such cases by integrating the AOD of the blue doublet component over the velocity range of the trough.

The 2002 and 2015b epochs include spectral coverage of \Lyb\ absorption from component 3. However, the red wing of the absorption trough is contaminated by strong absorption from molecular hydrogen.
We perform AOD integration on the blue half of the \Lyb\ trough, then double the value to obtain an estimation of the \hi\ column density from \Lyb.
The derived ionic column densities ($N_{ion}$) are given in Table 2.

\section{Photoionization Analysis}\label{secPhoto}
We used the spectral energy distribution (SED) continuum models
for the 2002 and 2015 epochs as described by \citet{Mehdipour18}, see their figure 2. 
To determine the photoionization structure of the outflowing gas, we used the same method described in \cite{Arav13}: using the above SED to run a grid of photoionization models with the spectral synthesis code Cloudy \citep[version c17.01][]{Ferland17}, by varying the total hydrogen column density (\Nh) and the hydrogen ionization parameter 
\begin{equation}\label{eqnUDef}
	\uh\equiv\frac{\qh}{4\pi R^2 \nh c}
	\label{eq:U}
\end{equation}
where \qh\ is the rate of hydrogen-ionizing photons from the central source, $R$ is the distance to the outflow from the central source, \nh\ is the hydrogen number density ($\ne\simeq 1.2\nh$ in highly ionized plasma) and $c$ is the speed of light.
We construct a phase plot depicting the solution by plotting the locus of models $(\uh,\Nh)$ that correctly predict the observed $N_\mathrm{ion}$ measurements (within $1\sigma$ uncertainty) for each ion.
These are shown in the top panel of \figref{figPhasePlotComp1AllEpoch} as contours spanned by colored bands representing the uncertainties.
We determine the best-fit model using chi-squared minimization.


\begin{table}
\caption{Column densities$^a$ for the NGC 7469 outflow components}\label{tabColDen}
\begin{tabular}{r@{$\,$}lr@{}c@{}lr@{}c@{}lr@{}c@{}l}
\hline
\multicolumn{2}{c}{Ion}	&\multicolumn{3}{c}{$v_3$}				&\multicolumn{3}{c}{$v_2$}								&\multicolumn{3}{c}{$v_1$}				\\
&			&\multicolumn{3}{c}{\small$[-2070,-1800]^b$}\hspace{-0.3cm}		&\multicolumn{3}{c}{\small$[-1470,-1370]^b$}\hspace{-0.3cm}				&\multicolumn{3}{c}{\small$[-700,-410]^b$}\hspace{-0.3cm}		\\\hline
 \multicolumn{11}{c}{Epoch 2002 \ \ \ \ \ \ relative UV flux 2.61$^c$}\\\hline
H&{\sc i}	&\color{blu}$>$&\color{blu}14.22&			&\color{blu}$>$&\color{blu}13.07&							&\color{blu}$>$&\color{blu}13.87&			\\
N&{\sc v}	&\color{blu}$>$&\color{blu}14.85&				&\color{red}$<$&\color{red}13.14&						&\color{blu}$>$&\color{blu}13.97&			\\
Si&{\sc iv}	&\color{red}$<$&\color{red}12.69&				&\color{red}$<$&\color{red}12.16&					&\color{red}$<$&\color{red}12.85&			\\
C&{\sc iv}	&\color{blu}$>$&\color{blu}14.56&				&\color{blu}$>$&\color{blu}13.37&						&&13.50&$^{+0.15}_{-0.15}$				\\
O&{\sc vi}	&\color{blu}$>$&\color{blu}14.98&$\,\!^d$				&\color{red}$<$&\color{red}13.77&				&\color{blu}$>$&\color{blu}14.92&			\\ \hline

 \multicolumn{11}{c}{Epoch 2004 \ \ \ \ \ \ relative UV flux 1.0}\\\hline
H&{\sc i}	&\color{blu}$>$&\color{blu}14.25&				&\color{blu}$>$&\color{blu}13.31&					&\color{blu}$>$&\color{blu}14.01&				\\
N&{\sc v}	&&14.67&$^{+0.37}_{-0.37}$					&\color{red}$<$&\color{red}12.62&					&\color{blu}$>$&\color{blu}14.11&				\\
Si&{\sc iv}	&\color{red}$<$&\color{red}12.57&				&\color{red}$<$&\color{red}12.78&				&\color{red}$<$&\color{red}13.33&				\\
C&{\sc iv}	&\color{blu}$>$&\color{blu}14.56&				&\color{blu}$>$&\color{blu}13.23&						&&13.79&$^{+0.1}_{-0.1}$					\\ \hline
	
 \multicolumn{11}{c}{Epoch 2010 \ \ \ \ \ \ relative UV flux 5.15}\\\hline
H&{\sc i}	&\color{blu}$>$&\color{blu}13.89&				&\color{blu}$>$&\color{blu}12.97&					&\color{blu}$>$&\color{blu}13.47&				\\
N&{\sc v}	&&14.73&$^{+0.19}_{-0.19}$					&\color{blu}$>$&\color{blu}13.09&						&\color{blu}$>$&\color{blu}13.62&				\\
Si&{\sc iv}	&\color{red}$<$&\color{red}12.72&				&\color{red}$<$&\color{red}12.26&				&\color{red}$<$&\color{red}12.37&				\\
C&{\sc iv}	&\color{blu}$>$&\color{blu}14.52&				&\color{blu}$>$&\color{blu}13.14&				&\color{red}$<$&\color{red}12.82&				\\ \hline

 \multicolumn{11}{c}{Epoch 2015a \ \ \ \ \ \ relative UV flux 2.26}\\\hline
H&{\sc i}	&\color{blu}$>$&\color{blu}14.21&				&\color{blu}$>$&\color{blu}13.06&				&\color{blu}$>$&\color{blu}13.70&				\\
N&{\sc v}	&&14.93&$^{+0.65}_{-0.65}$					&\color{red}$<$&\color{red}13.35&					&\color{blu}$>$&\color{blu}13.90&				\\
Si&{\sc iv}	&\color{red}$<$&\color{red}13.02&				&\color{red}$<$&\color{red}13.16&				&\color{red}$<$&\color{red}12.59&				\\
C&{\sc iv}	&\color{blu}$>$&\color{blu}14.89&				&\color{blu}$>$&\color{blu}12.99&						&\color{blu}$>$&\color{blu}13.26&		\\ \hline

 \multicolumn{11}{c}{Epoch 2015b \ \ \ \ \ \ relative UV flux 1.94}\\\hline
H&{\sc i}	&\color{blu}$>$&\color{blu}14.12&				&\color{blu}$>$&\color{blu}13.01&							&\color{blu}$>$&\color{blu}13.65&				\\
N&{\sc v}	&&14.95&$^{+0.30}_{-0.30}$					&\color{red}$<$&\color{red}13.39&						&\color{blu}$>$&\color{blu}13.84&				\\
Si&{\sc iv}	&&12.95&$^{+0.20}_{-0.20}$					&\color{red}$<$&\color{red}12.86&						&\color{red}$<$&\color{red}12.39&				\\
C&{\sc iv}	&\color{blu}$>$&\color{blu}14.81&				&\color{blu}$>$&\color{blu}13.16&			&\color{blu}$>$&\color{blu}13.34&					\\
O&{\sc vi}	&\color{blu}$>$&\color{blu}15.08&				&\color{red}$<$&\color{red}13.98&			&\color{blu}$>$&\color{blu}14.88&				\\ \hline
 \multicolumn{11}{l}{$\,^a$Table values are log column densities  ($\mathrm{cm}^{-2}$)} \\ 
 \multicolumn{11}{l}{ \ \  Lower limit   are shown in {\color{blu} blue}, upper limits   in {\color{red} red}.}\\
 \multicolumn{11}{l}{ \ \  Measurements (and errors) are shown in black.}\\\multicolumn{11}{l}{$\,^b$Integration limits in km~$\mathrm{s}^{-1}$.}\\
 \multicolumn{11}{l}{$\,^c$flux at 1170~\AA\ relative to the 2004 flux}  \\ 
 \multicolumn{11}{l}{ \ \     (2004 flux at 1170~\AA\ = $1.20 \times \rm 10^{-14}~erg~cm^{-2}~s^{-1}~\AA^{-1}$)}\\
\multicolumn{11}{l}{$\,^d$The \ovi\ absorption trough only exists in Epoch 2002}\\
\multicolumn{11}{l}{\hspace{0.21cm}  and Epoch 2015b. See Section 2 for more details.}\\
\end{tabular}
\vspace{-0.7cm}
\end{table}

\subsection{Component 1}

Component 1 has a substantial optical depth and exhibits the strongest variability of all three components between multiple epochs (see \figref{figEpochNormLyaAllComp}). We begin by constructing and discussing photoionization solutions, first for the 2015 data and then for all epochs. We then use time-dependent photoionization analysis to model the troughs variability, which leads to a distance determination for this component from the central source

\subsubsection{Consistent photoionization solutions for all epochs}

The portion of the SED responsible for the photoionization equilibrium of the species we observe with COS (\hi, \civ, \nv, \ovi\ and \siiv) is between 13.6 eV to several hundred eV. 
Across this energy range, the ratio of the 2015 to the 2002 SED changed by a maximum of $\pm15$\% \citep[see figure 2 in][]{Mehdipour18}.
Thus, we  make the simple and restrictive assumption that the shape of the NGC 7469 Spectral SED is the same in all observed epochs from 2002--2015.  Therefore, the changes in integrated flux and \uh\ are equal to changes in the  measured UV flux (see Table 2). Since \Lya\ is a singlet, we methodically treat its individual AOD $N_{ion}$ measurements as lower limits (see Table 2).  
However, component 1 show trough variations that strongly correlate with the measured UV flux (see 
\figref{figEpochNormLyaAllComp} and Table 2). Quantitatively,
from Table 2 we infer that for the largest flux variation, a factor of 5.2 ($\Delta\log(\uh)=0.7$)  between 2004 and 2010,
the AOD derived N(\hi) changed by a factor of 3.5. With no saturation, and assuming photoionization equilibrium, the 2004 N(\hi) should have been $\sim7$ times higher than in 2010 
(where we assumed  $\log({\uh}_{2004})=-1.3$ and $\log({\uh}_{2010})=-0.6$, see below). 

Assuming that a) the real change in N(\hi) between the two epochs is indeed a factor of 7; and b) that the \Lya\ trough had the same covering factor in 2004 and 2010, we can perform a covering factor analysis on the two troughs \citep[see section 3 of ][for the full formalism, including velocity dependence]{Arav05}. The results are that the real N(\hi) is only a few percent larger than the AOD value deduced from the shallower 2010 trough, but 100\% larger than the AOD value deduced from the deeper 2004 trough. This result is consistent with the findings that
when troughs from the same ionic species are saturated, the deeper trough is more saturated than the shallower one 
\citep{Arav18}. We therefore use the 2010 N(\hi) reported value as an actual measurement, and consider the  deeper 2004 trough to be mildly saturated, where its N(\hi) is two times the lower limit reported in Table 2. 

Independent support for the mild-saturation scenario comes from the \Lyb\ observations in the 2002 and 2015b epochs.
The center and entire blue wing of the \Lyb\ trough of component 1 ($-700<v<-500~\kms$) are heavily contaminated with galactic absorption.  However, at roughly $-500\kms$ the galactic absorption disappears, and at that velocity, for component 1, an upper limit of $\tau$(\Lyb)=0.05 is derived from the data for both epochs.  At the same velocity (equivalent to 1233.4\AA\ observed wavelength in figure 2), 
$\tau$(\Lya)=0.32 and 0.25 for the 2002 and 2015b epochs, respectively. For a case of no saturation (i.e., an optically thin absorber) $\tau$(\Lya)/$\tau$(\Lyb)=6.2  (based on the ratio of the product of their oscillator-strength and wavelength). Therefore, the observed ratios between the measured $\tau$(\Lya) and upper limit on $\tau$(\Lyb) at $-500\kms$ (6.4 and 5, for the 2002 and 2015b epochs, respectively) are consistent with the expected value for an unsaturated absorber.     

We note that the \civ\ and \nv\ troughs respond similarly to changes in the UV flux, and therefore the nominal lower limits on their $N_{ion}$ are close to their actual values. 

\begin{figure}
\vspace{-.5cm}
\includegraphics[width=\columnwidth,angle=90]{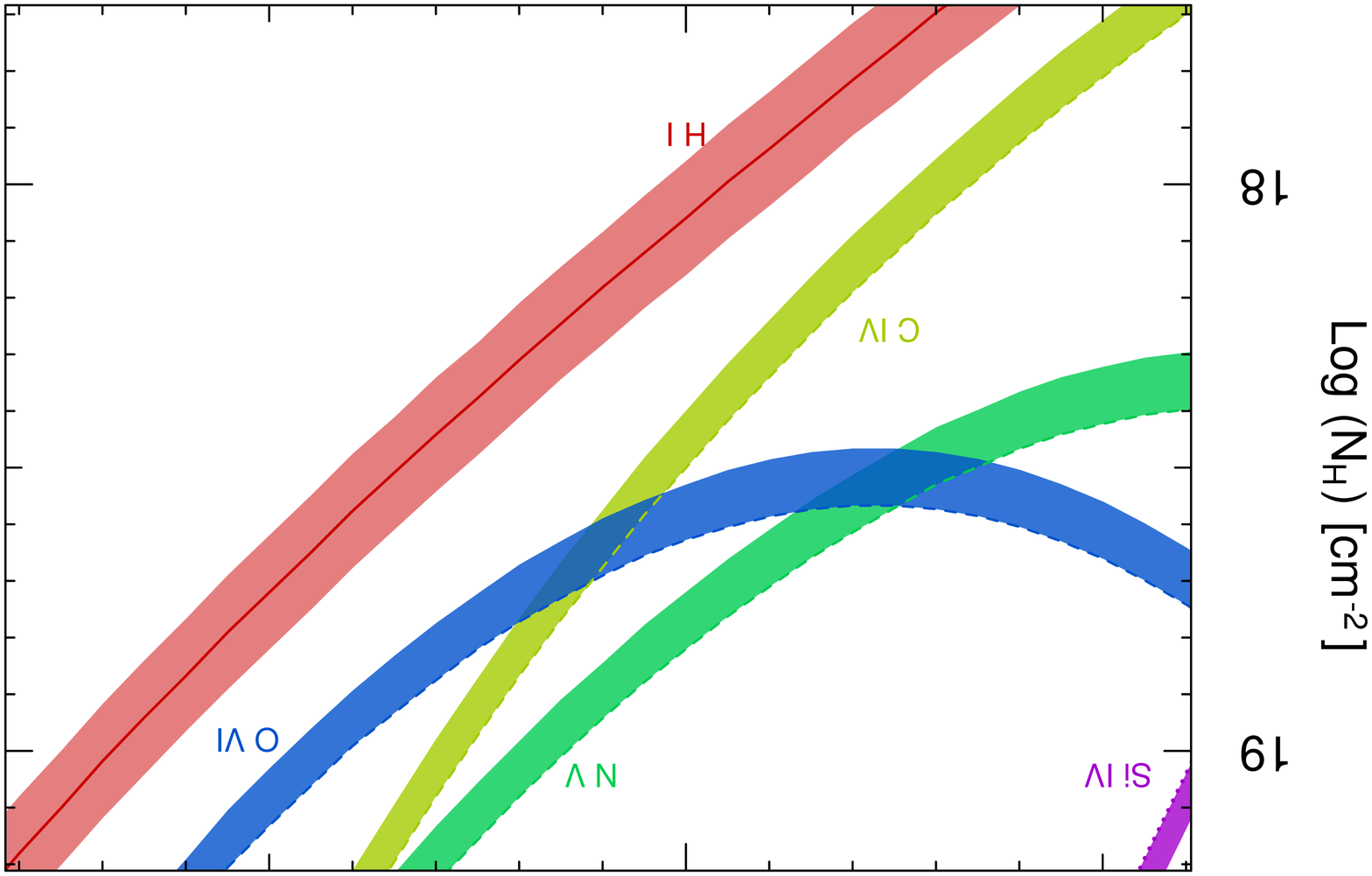}
\vspace{-3.5cm}*
\includegraphics[width=\columnwidth,angle=90]{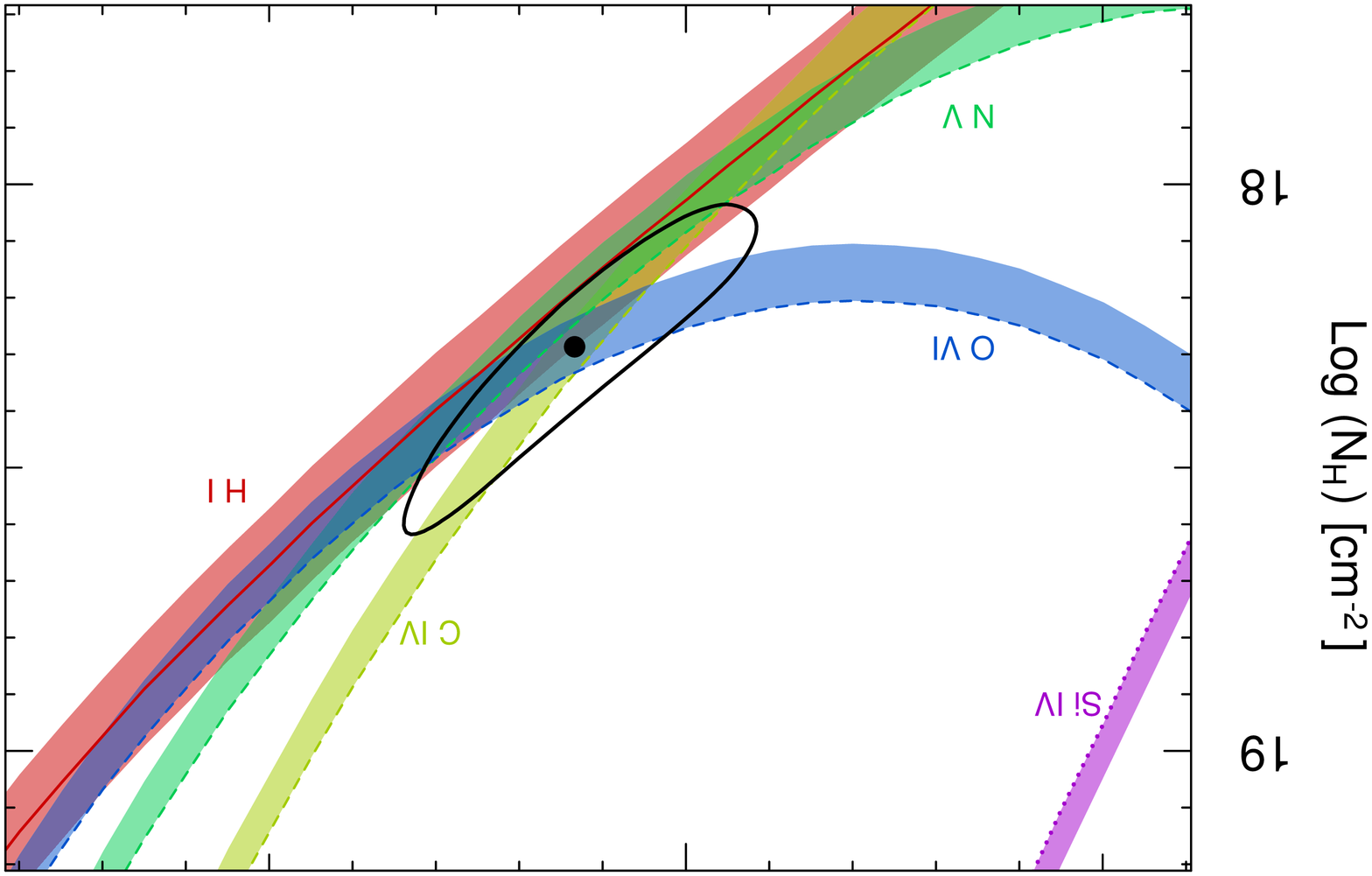}
\vspace{-3.5cm}*
\includegraphics[width=\columnwidth,angle=90]{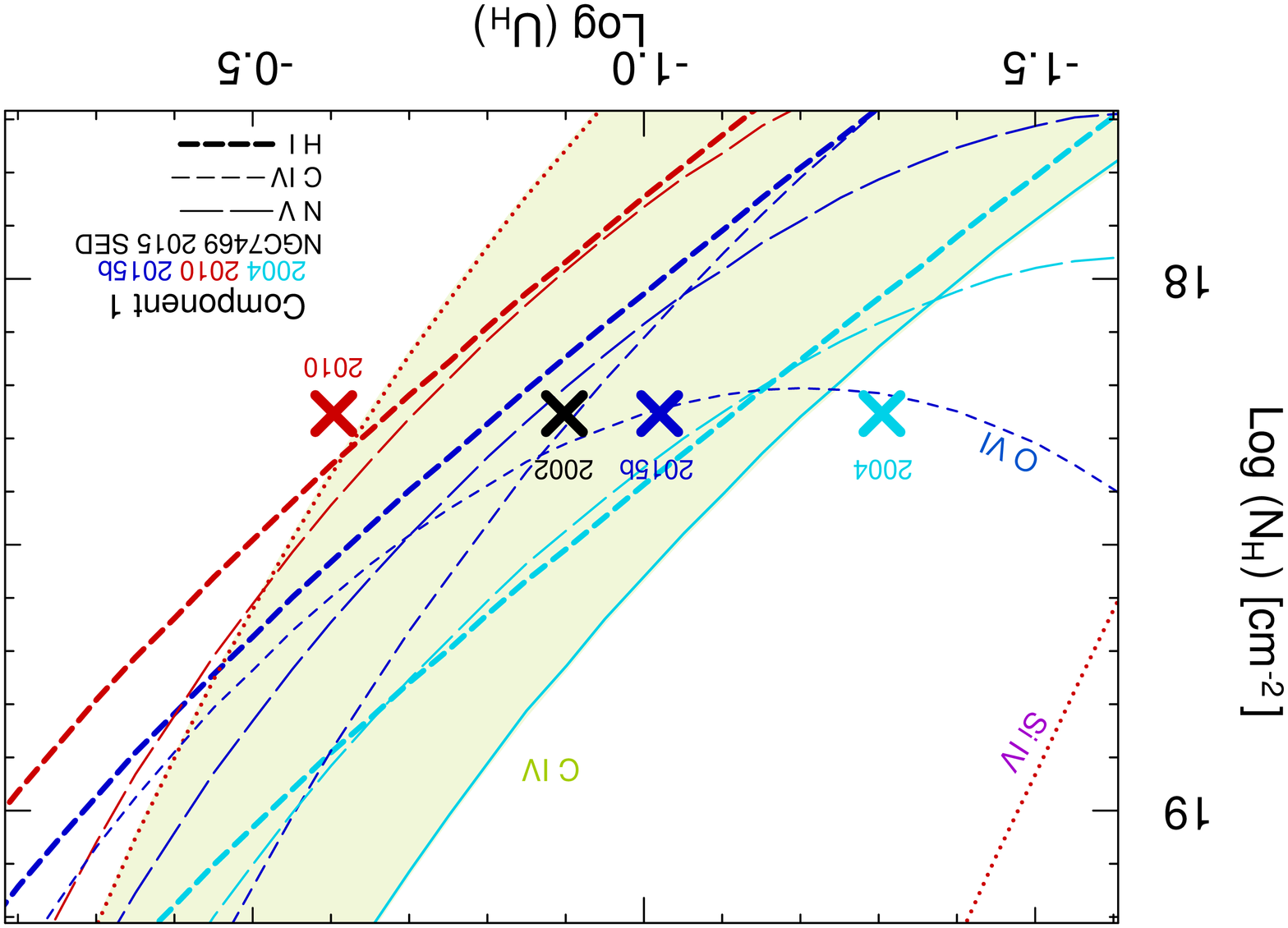}
\vspace{-2.5cm}
\caption{\textbf{Top:} Photoionization phase plot showing the constraints for component 1 of epoch 2015b, where we use the 2015 NGC~7469 SED and proto-solar metallicity.
Solid lines and associated colored bands represent the locus of \uh, \Nh\ models, which predict the measured $N_\mathrm{ion}$, and their $1 \sigma$ uncertainties;  dashed lines represent $N_\mathrm{ion}$ lower limits  that permit a solution in the phase-space above them;  and dotted lines represent $N_\mathrm{ion}$ upper limits  that permit a solution in the phase-space below the line. Since the CNO lower limit bands do not intersect with the \hi\ band, it is clear that there is no viable photoionization solution for this case.
\textbf{Middle:} Same $N_\mathrm{ion}$ constraints and same SED as in the top panel, but assuming  \textbf{super-solar metallicity} as described in section 4.1.1. The viable physical 
solution and its 1-$\sigma$ error are shown by the black dot and surrounding black oval, respectively. 
\textbf{Bottom:}
Phase plot showing the photoionization constraints for  component 1 in three different epochs,
using the same  SED and abundances used in the middle panel.
Similar presentation to the other panels, where for clarity’s sake we do not show
the errors on individual measurements, (the errors are similar to the ones shown in the middle panel). 
Most of the \civ, \nv\ and \Lya\ constraints are lower limits. To help distinguish between them, we assign different length dashed lines for these lower limit constraints.
The region between the three \civ\ constraints is shaded to help guide the eye. The `X' symbols show the location of the derived photoionization solutions for the different epochs where \Nh\ is kept constant and the 
difference in \uh\ is equal to the difference in the measured UV flux for each epoch.}
\label{figPhasePlotComp1AllEpoch}
\end{figure}

The top panel of figure \ref{figPhasePlotComp1AllEpoch} shows the photoioinization phase plot for the 2015b epoch, where we used the 2015 SED 
and assumed the proto-solar abundances given by \cite{Lodders09}. For \civ, \nv\ and \ovi, the $N_{ion}$ constraints are all lower limits, requiring the photoioinization solution to be inside or above their colored bands. Since we established that the N(\hi) constraint is a measurement,  the photoioinization solution has to be inside the red band. 
It is evident that there is no region on this plot that satisfies both the hydrogen and CNO constraints.

A physically plausible way to alleviate this issue is to invoke Super Solar Metallicity (SSM). As shown in dedicated studies, AGN outflows exhibit SSM.
In the case of Mrk 279 (a Seyfert I with similar luminosity to NGC 7469) we found the following CNO abundances relative to solar (using the standard logarithmic notation) [C/H]=0.2--0.5, [O/H]=0--0.3 and [N/H]=0.4--0.7 \citep{Arav07}.
For the outflow observed in the Hubble-deep-field-south target QSO J2233--606, we found [C/H],   [O/H] $\approx 0.5-0.9$ and
[N/H] $\approx 1.1-1.3$ \citep{Gabel06}.  The SSM for the outflows in both objects is consistent with enhanced nitrogen production expected from secondary
nucleosynthesis processes \citep{Gabel06}, where [N/H]$\simeq$2[C/H]$\simeq$2[O/H]$\simeq$2[Si/H] \citep[see model M5a of][which is used for obtaining abundances with different metallicities than proto-solar in Cloudy]{Hamann93}.

In constructing a viable SSM photoionization model, we attempt to 
restrict the number of degrees of freedom (associated with SSM and otherwise) as much as possible. First, we assume that the chosen SSM must apply to all epochs. Second, we use metallicity enhancement that is both consistent with the Mrk 279 results and adheres to the enhanced nitrogen production expected from secondary nucleosynthesis processes described above.  Thus, we use 
[C/H]=[O/H]=[Si/H]=0.35 and [N/H]=0.70, which  
yields a viable solution for the 2010 epoch constraints.

In the middle panel of figure \ref{figPhasePlotComp1AllEpoch} we show the photoioinization phase plot for the 2015b epoch, where we use the 2015 SED 
and assumed the SSM abundances values given above. The $N_\mathrm{ion}$ constraints are the same but their position on the \uh--\Nh\ plane is lower by roughly the inverse of their scaled abundances.  This is because for an enheced abundance of a given element the \Nh\ needed to satisfy its $N_\mathrm{ion}$ is smaller. The best fitting solution and its 1-$\sigma$ error are shown by the black dot and surrounding black oval, respectively.

Changing the abundances yields a solution for the 2015b epoch mainly due to the introduction of two new parameters (the coupled abundances for C, O and Si, and the abundance of N). It is the fits to the other epochs that makes this model physically robust. The models for the other epochs have no additional free parameters.  Similar to our model for the NGC 5548 outflow \citep{Arav15}, we require that: a) the \uh\ of the solution for each different epoch will differ by the ratio of the UV flux of the said epoch to the UV flux of the 2010 epoch (for which the abundances were determined to allow for a viable solution); and b)  \Nh\ is constant for all epochs. The last requirement follows our assertion below, that it is unlikely that significant amount of material will enter or exit the line of sight between 2002 and 2015 (see section 6).

The model for all 4 epochs is shown in the bottom panel of Figure \ref{figPhasePlotComp1AllEpoch}.
For clarity's sake, we show constraints from only 3 epochs: those with the lowest and highest UV fluxes, 2004 and 2010, respectively,  and 2015b, which has the highest S/N data and is the most recent.  We also do not show the 1$\sigma$ error bars associated with the different constraints, but they are similar to the width of the colored ribbons on the top panel. Each of the solutions fit all the $N_{ion}$ constraints roughly within the error. 
We note that: \\
a) the \siiv\ constraint for all epochs is an upper limit which is trivially satisfied in all epochs as the solution needs to be below the upper limit curve.  \\
b) we implicitly assume that the \civ\ and \nv\ lower limits $N_{ion}$ reported in Table 2 are actual measurements as we ascribe their different values to our modeled photoionization changes.\\ 
Thus, this restrictive and simple model, which is based on a fixed total column-density absorber reacting to changes in ionizing illumination, matches the
measured constraints spanning 13 years.

We note that even assuming the same SSM, the \uh--\Nh\ solution for the 2015b epoch in the bottom panel is slightly different than the one in the middle panel. That is because the middle panel solution was optimized to only the 2015b 
$N_\mathrm{ion}$. On the bottom panel all the solutions are fixed by the 2010 solution (see above) and therefore the \uh--\Nh\ solution for 2015b in the lower panel is not the same as in the middle panel.  An important aspect of the 4 epochs solution is that the \uh--\Nh\ solution for 2015b in the bottom paneel is within the error ellipse of the pure 2015b solution.

\begin{figure}
\includegraphics[height=\columnwidth,angle=90]{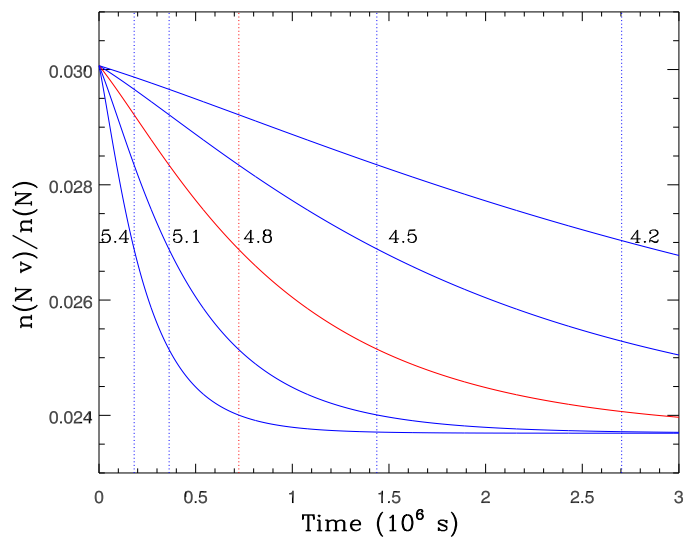} 
\caption{Time-dependent photoionization solutions:
starting from the photoionization equilibrium for the 2015b epoch (shown in the bottom panel of figure \ref{figPhasePlotComp1AllEpoch}), we assume a step-function flux increase of 13\% (see text), 
and track the changes in the relative fraction of \nv\ for a range of \ne. The log(\ne) of each solution labels each curve. The shortest time-scale where we measured definitive changes in the \nv\ trough between two epochs was 700,000 seconds. In red we mark the \nv\ fraction curve that shows a 50\% change between the initial and final values after 700,000 seconds.
}
\label{figTime_dep_PI}
\end{figure}

\subsubsection{Distance determination from trough variability}

\begin{table}
\caption[]{Time of 2015 Visits and Flux}
\label{Table2015obs}
\begin{tabular}{c c c c c c c}
\hline\hline
Obs & Time & $f_{1170}$  \\
    & (MJD)$^{a}$& ($\rm 10^{-14}~erg~cm^{-2}~s^{-1}~\AA^{-1}$) \\
\hline
V01-2015 & 185.84 & $2.71 \pm 0.01$ \\
V02-2015 & 350.62 & $1.94 \pm 0.01$ \\
V03-2015 & 371.43 & $2.32 \pm 0.01$ \\
V05-2015 & 379.38 & $2.63 \pm 0.01$ \\
V06-2015 & 381.30 & $2.46 \pm 0.01$ \\
V07-2015 & 382.95 & $2.20 \pm 0.01$ \\
V09-2015 & 385.20 & $2.44 \pm 0.01$ \\
\hline
$^{a}$MJD-57,000 
\end{tabular}
\end{table}

Component 1 exhibits variations in Ly$\alpha$, \civ\ and \nv\ absorption strengths during the visits of the 2015 epochs. As detailed above, it is likely that these trough variations are the response of the absorber to changes in the ionizing flux.
Tracking changes in column density of
a given ion between different epochs along with flux monitoring
can lead to estimates of \ne. Then using equation (1), the distance $R$ can be determined \citep[e.g.,][]{Gabel05b}.

The timescale for photonionization changes is inversely proportional to $n_e$. Therefore, for a given 
\ne\ there is a characteristic time-scale where flux changes on longer time periods will change the ionization fraction of an ionic species appreciably, while changes on shorter periods will not. 
Our program was designed to help identify this time-scale by methodically sampling a wide range of $\Delta t$ between the different visits.  The 7 visits of the main monitoring program  had  $\Delta t$ of 165, 21, 8, 1.9, 1.7 and 2.3 days between consecutive visits starting with visit 1 (see Table 3). We exclude visits 4 and 8 (see Table 1), which were carried out only with the G130M grating and had a different central wavelength setting.  

The shortest timescale during which changes in the \nv\ absorption troughs were observed occured between visits 3 and 5, corresponding to $\Delta t = 8.0$ days or $\sim$700,000 seconds. The flux at 1170~\AA\ increased between these visits by 13\%. Similarly, the \civ\ trough varied between visits 7 and 9, a separation of $\Delta t = 2.25$ days, with a  10\% flux change.

In order to constrain \ne\ for outflow component 1, we numerically solve the time-dependent photonionization  differential-equation set   \citep[see equation (6) in][]{Arav12}. In what follows, we describe the \nv\ based analysis.
Our starting point is the  photoinization equilibrium for component 1, at the 2015b epoch, shown in figure \ref{figPhasePlotComp1AllEpoch}. This Cloudy solution yields the population fraction  for each nitrogen ion compared to nitrogen as a whole, as well as the recombination coefficients for all ions. We then make the simplified assumption that the observed 13\% flux change is in the form of a step function that occurs immediately after the initial state. Following these initial conditions, in figure \ref{figTime_dep_PI}, we show how the \nv\ fraction changes over time. Different curves correspond to different log(\ne) values. The dotted vertical lines show the value of $t_{50\%}$, which is defined as the time scale where a 50\% change in the \nv\ fraction was achieved. That is, $t_{50\%}$ is defined by: \\
n[\nv]($t_{50\%})$=[n[\nv]($t=0)$--n[\nv]($t=\infty$)]/2.

The red curve with log(\ne)=4.8 has $t_{50\%}$ of 8 days (700,000 seconds), which matches the shortest 
$\Delta t$ where we observed changes in the \nv\ trough. The log(\ne)=4.5 curve does not show
a large enough change in n[\nv] over 8 days to explain the observed trough variability. 
We do not detect changes in the \nv\ trough between epochs 5-6, 6-7 and 7-8 (all with $\Delta t \simeq 2$ days), where flux changes are around 10\% between neighboring epochs. Similar analysis shows that 
for log(\ne)=5.1,  a measurable change should have been detected in the \nv\ trough in 4 days (350,000 sec). However, between epochs 5 and 7 ($\Delta t=3.6$ days) the flux changed monotonically by 
16\%, but there are no definitive changes in the \nv\ trough.  From this behavior, we conclude that 
log$(\ne)<5.1$ cm$^{-3}$.
Thus, the \nv\ constraints yield log$(\ne)=4.8\pm0.3$ cm$^{-3}$ for outflow component 1.
A similar analysis of the \civ\ trough variation yields a consistent constraint of log$(\ne)>4.7 \pm0.3$ cm$^{-3}$.

With the above  $n_e$ determination for component 1, we can  solve equation~(\ref{eq:U}) for 
$R$. For the \uh\ value of component 1, $n_e=1.2\nh$.  The ionizing photon rate ($Q_H$) is calculated by integrating the  2015 SED  for energies above 1 Ryd and normalized to the flux at 1170 \AA, and we obtain $Q_H=5.5\times 10^{53}$~s$^{-1}$. With these parameters, equation~(\ref{eq:U}) yields R=$6^{+2.5}_{-1.5}$ pc.

\subsection{Component 2}

\subsubsection{Photoionization Solution for the 2015 COS data}

The absorption troughs of component 2 are shallow (residual intensity $\gtorder0.8$).  Therefore, the relative errors for a given signal-to-noise-ratio (SNR) are much larger than for the other components, which makes the analysis results more uncertain.  Using the SSM described in section 4.1.1 and the 2015 SED, a rough photoionization solution for the 2015b epoch is:
log(\Nh)=17.3 (cm$^{-2}$) and log(\uh)=-1.5.

\subsubsection{Variability analysis}

From figure \figref{figEpochNormLyaAllComp} we note that the \Lya\ troughs of four epochs are consistent with no trough changes, while the 2004 epoch seems deeper. 
Qualitatively, we expect the 2004 trough to be deeper if the absorber reacts to changes in the ionizing flux. This is because the UV flux (and therefore \uh) of the 2004 epoch is the smallest and therefore its N(\hi) should be the largest (assuming no changes in total \Nh\ during the various epochs).  However, the changes in the UV flux between the 2015b epoch and the 2004 epoch (where the trough is somewhat deeper), are smaller than those between the 2015b epoch and the 2010 epoch (where there are no trough changes detected).  Since the 2015b and 2010 data sets have higher SNR than the 2004 epoch it is more probable that the trough did not change between all the epochs, and that the 2004 changes are due to the low SNR of that epoch.  We note that the 2004 epoch shows another absorption feature that is not seen in any other epoch at 1229.3\AA\ observed wavelength, very close to the position of component 2. Assuming there is no trough variability between the 2002 and 2015 epochs, the distance of component 2 at least an order of magnitude farther from the central source than component 1, or $\sim60$~pc.

\subsection{Component 3}
Component 3 is the deepest one in the  outflow and shows a complicated pattern of variability over 13 years.  We begin by constructing and discussing photoionization solutions for the 2015 data, first with proto-solar abundances and then assuming the same abundances we used for component 1. Following that, we present the unusual variability of this component and 
attempt to model this variability with two ionization phases. We then discuss the successes and limitation of this model.

\subsubsection{Photoionization Solution for the 2015 COS data}
\begin{figure}
\includegraphics[width=\columnwidth,angle=90]{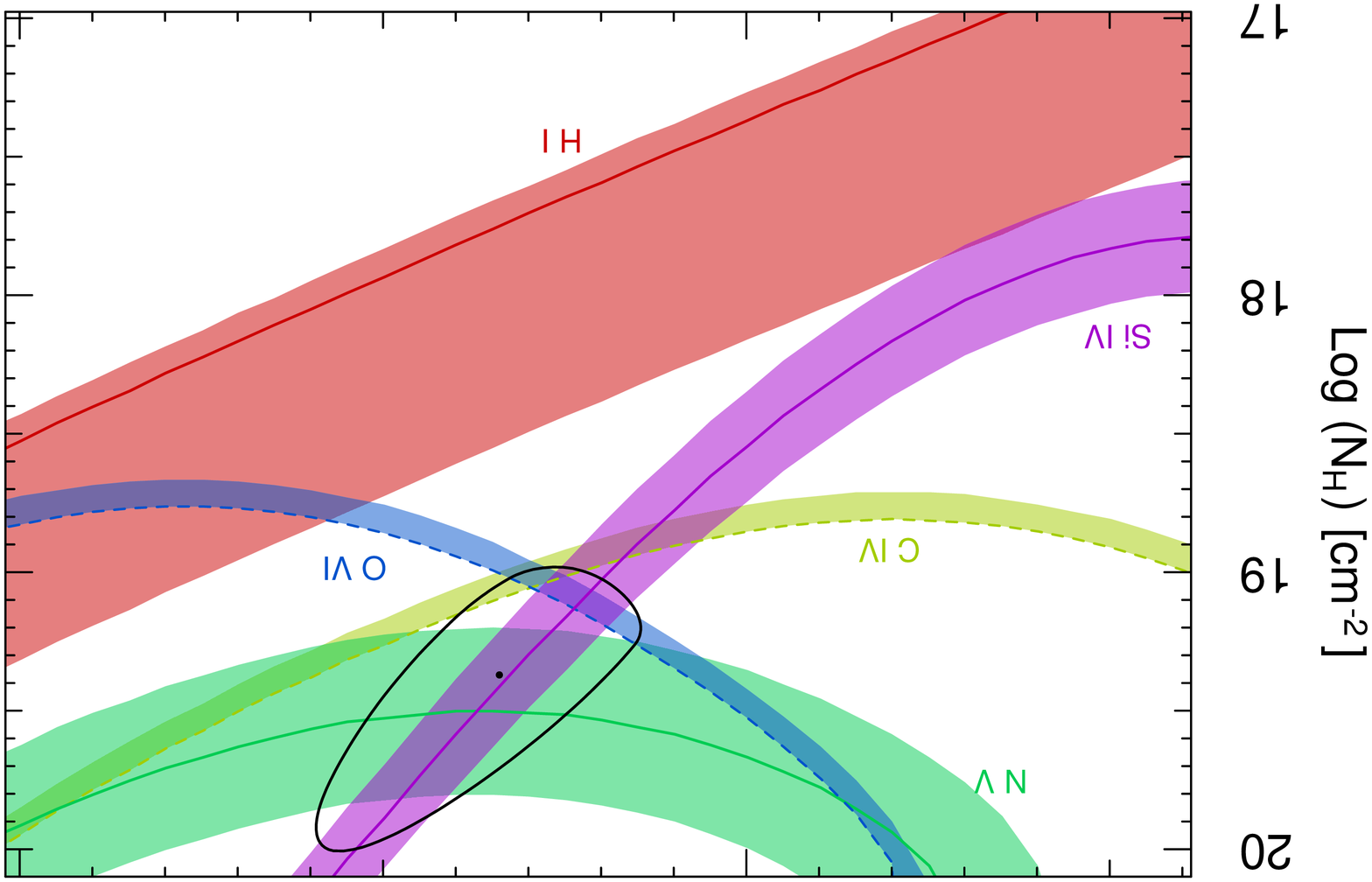}
\vspace{-3.5cm}*
\includegraphics[width=\columnwidth,angle=90]{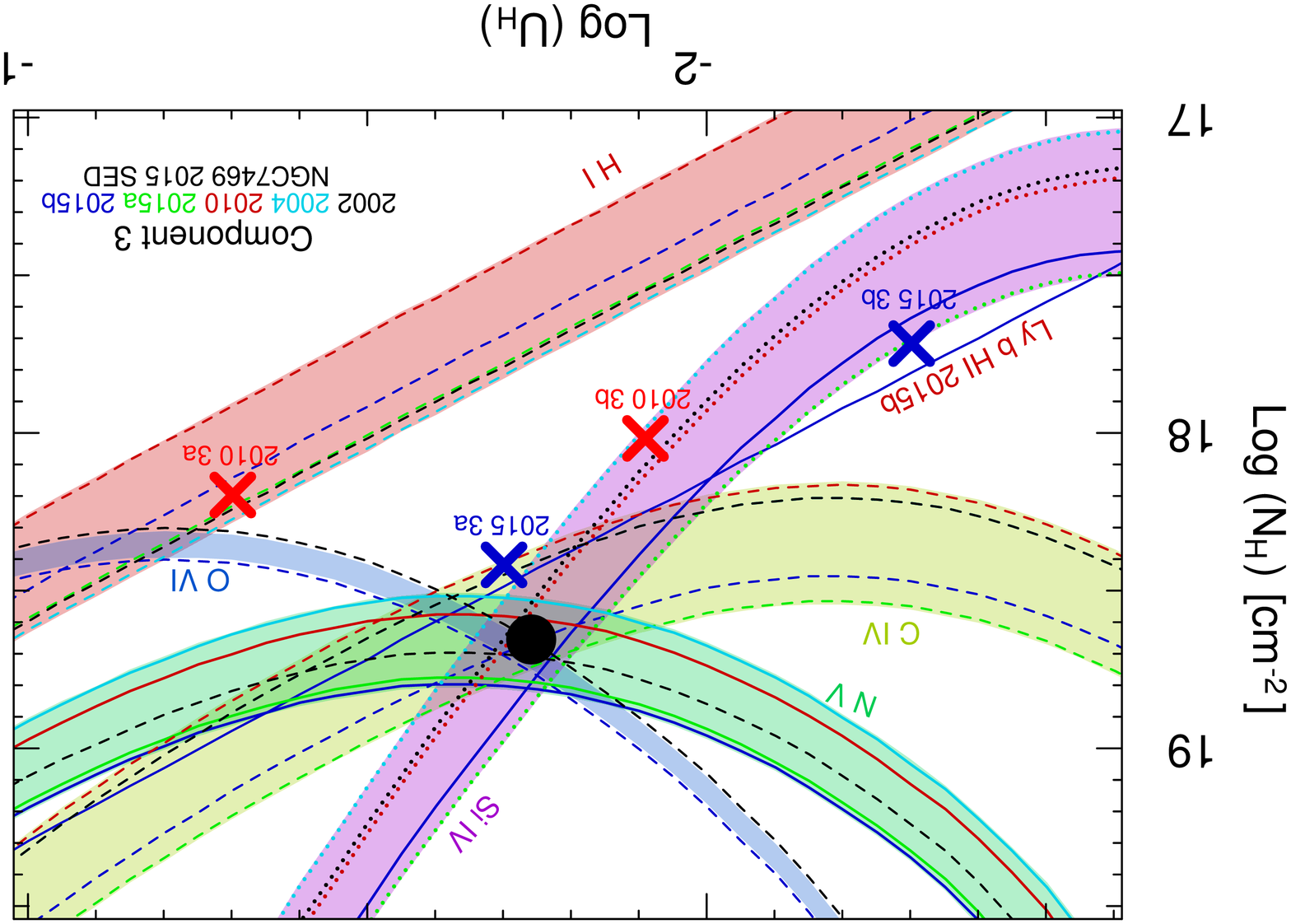}
\vspace{-2cm}
\caption{\textbf{Top panel:} Photoionization phase plot showing the ionization solution for component 3 of epoch 2015b.
We use the 2015 NGC~7469 SED and proto-solar metallicity.
Solid lines and associated colored bands represent the locus of \uh, \Nh\ models, which predict the measured $N_\mathrm{ion}$, and their $1 \sigma$ uncertainties, while the dashed line is the lower limit on the \ovi\ column-density that permits the phase-space above it.
The black dot is the best $\chi^2$ solution and is surrounded by a 1$\sigma$ $\chi^2$ black contour.
\textbf{Bottom panel:} Component 3 multi-epoch photoionization phase plot,  using the same enhanced metallicity we used for component 1.
Measurements are shown as solid lines, lower limits as dashed lines, and upper limits as dotted lines. The measurements and limits are colored according to epoch of observation.
The colored bands envelop measurements and limits for the same ion. For clarity's sake,  we do not show the errors on individual measurements (which are shown in the top panel 
 for the 2015b epoch and are representative to all epochs). The nominal solution that satisfies all the $N_{ion}$ constraints for the 2015b epoch is shown by the black circle.
In section 4.3.2 we describe a 2-phase model for fitting the variability seen in component 3.
We show the \uh\ and  \Nh\ positions of these phases (3a and 3b) for the  2010 and 2015 epochs with red and blue  "X" symbols. 
}
\label{figPhasePlotComp3AllEpoch}
\end{figure}

In the top panel of \figref{figPhasePlotComp3AllEpoch}, we show  the phase plot for the 2015b epoch
assuming proto-solar abundances.
The \siiv\ and \nv\ measurements for that epoch offer the strictest constraints on the photoionization solution, and the lower limits on the \ovi\ and \civ\ column densities exclude the lower portion of the \siiv\ contour.
Together, these constraints locate the solution at $\log \uh=-1.67\pm 0.2$ and $\log \Nh=19.39\pm 0.5 \mathrm{~cm}^{-2}$.
The \hi\ ionic column density deduced from \Lya\ is treated as a lower limit since the presence of absorption from the weaker line of \Lyb\ requires a larger \hi\ column density.
The \Lyb\ constraint is represented by the upper error on the \hi\ shaded contour in \figref{figPhasePlotComp3AllEpoch}, which remains a factor of six (in \Nh) below the solution.
Thus, based solely on the 2015b data, the proto-solar abundances suggest that \Lya\ is saturated by a factor of 40 as the vertical separation between the solution and the position of the \Lya-based \hi\ line is 1.6 dex. 

However, data from the other epochs show an anti-correlation between the \hi\ $N_{ion}$ (deduced from \Lya)
and the UV flux as expected from photoionization changes. Such a behavior cannot occur if the saturation level is 40, as suggested by the  proto-solar abundances. With such a high saturation, the shape of the trough is almost entirely due to velocity dependent covering-factor, 
 and trough variability due to real \hi\ column-density changes are negligible \citep[e.g.,][]{Arav08}. Using the same abundances that gave a good fit to component 1  alleviates this issue considerably. The photoionization phase diagram based on the same SED and $N_{ion}$ measurement, but using the  [C/H]=[O/H]=[Si/H]=0.35 and [N/H]=0.70 abundances, is shown in the bottom panel of 
\figref{figPhasePlotComp3AllEpoch}. It is evident that this model gives an excellent fit for all the ionic constraints as the nominal solution  at  $\log \uh=-1.7$ and $\log \Nh=18.6 \mathrm{~cm}^{-2}$ (shown by the filled black circle), is consistent with all the $N_{ion}$ constraints, to about $1\sigma$, including the \hi\ \Lyb\ constraint. We note that the different abundances of this model were not akin to introducing two unrestricted degrees of freedom, (which would have made a good fit trivial).  Rather, we used the exact same (physically-motivated) abundances that gave a good fit to component 1. In contrast, as we'll show in the next section, attempting a quantitative variability modeling for component 3 suggests a change in the physical picture of the absorber.

\subsubsection{Modeling Trough Variability}
\begin{figure}
\includegraphics[width=\columnwidth]{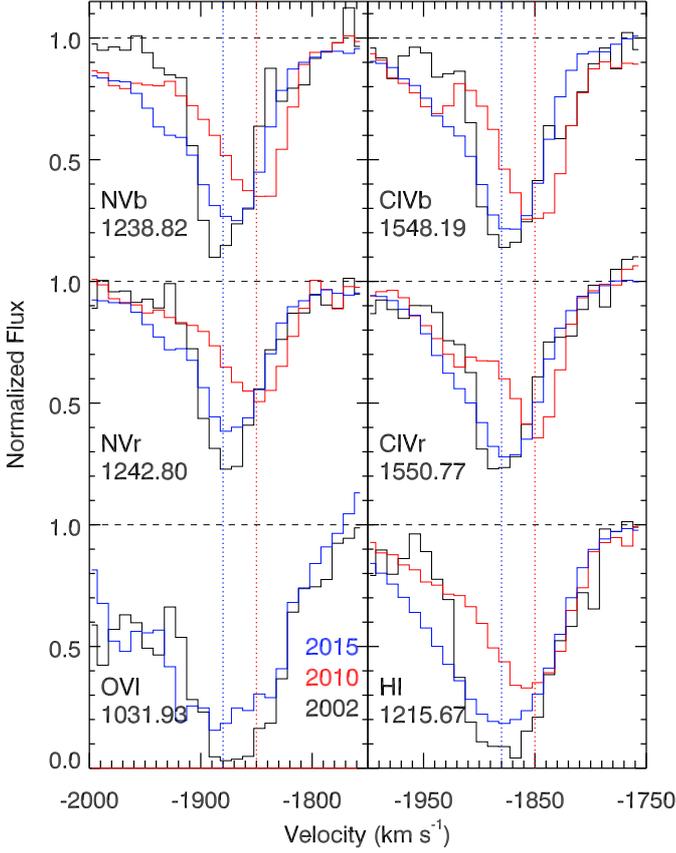}
\caption{Component 3 variations: normalized spectrum showing the velocity-shift anomaly in the 2010 epoch.
We illustrate this anomaly in outflow component 3 by showing vertical dotted lines through the centroid of the absorption troughs in the 2002 and 2015 epochs (blue) and during the 2010 epoch (red). We note that the overall depth of component 3 is lowest in the 2010 epoch in all of the observed troughs.}
\label{figComp3AllEpoch}
\end{figure}

Component 3 exhibits complex variations in the shape of its absorption troughs during the five spectroscopic epochs.
In particular, the 2010 observations show that the velocity centroid of the trough shifted by +30~\kms\ compared with the previous epochs. Initially, we thought that the trough velocity decelerated \citep[like is seen for one component in NGC 3783,][]{Gabel03}.
However,  in the 2015 observation  the velocity centroid of component 3 shifted back to its position in the 2002 and 2004 epochs.
Component 3 exhibits this shift in each ion observed in 2010 (see \figref{figComp3AllEpoch}).
Components 1 and 2 do not show this behavior, confirming that the shift is not due to a wavelength calibration issue.

\begin{figure}
\includegraphics[width=\columnwidth]{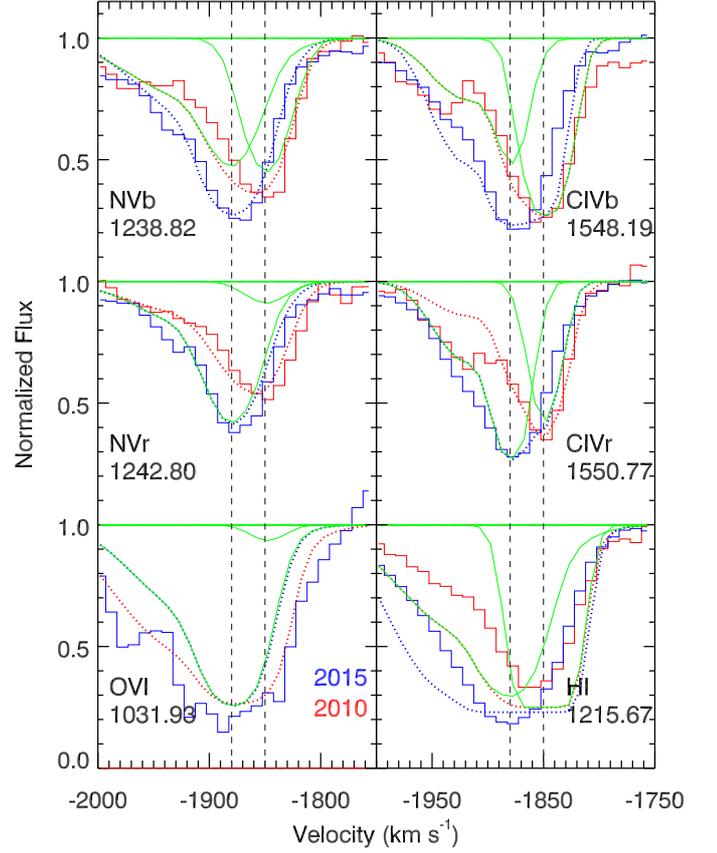}
\caption{Model fit for troughs variation in component 3 (see \S~4.3.2). The normalized data is shown in red for the 2010 epoch and blue for the 2015 one. The full model for each trough is shown by the dotted lines of the same color. In green we show the contribution of component 3a (left curve) and 3b (right curve). For the blue doublet components (left panels) we show the contribution of the 2 phases to the 2010 model, and for the red doublet components (right panels) we show the contribution of the 2 phases to the 2015 model.}
\label{figComp3model}
\end{figure}


The key for understanding this unusual behavior of component 3 comes from taking into account the UV flux level in the different epochs.
As can be seen in Table 2, the UV flux during the 2010 observations was the highest of all epochs, roughly twice  the flux of the 2002 and 2015 epochs, and 5 times higher than the 2004 epoch. We therefore attempt to model the trough variations using the variation in the photoionization equilibrium induced by the changes in UV flux between the epochs. 
We note that such a model can, in principle, match most of the trough variations. However, there must be other effects as well. For example, while the high velocity wing changes uniformly between the 2010 and 2005 epochs in all of the troughs, the 2002 epoch is deeper close to the centroid and shallower at the high velocity wing.  This 3-epoch comparative phenomenology cannot be explained by pure photoionization changes.

A model based on a single ionization phase for component 3 is inadequate for this purpose, as it predicts a uniform change in a trough's depth across the entire profile.
We therefore attempt to model component 3 with two ionization phases whose velocity centroid is shifted by 30 \kms\ relative to each other. Our approach is to build the simplest possible two-phase model that can fit the major empirical behavior seen in the data.  Following this principle, we construct a model that has two kinematic components (3a and 3b) that differ by a constant $\Delta$\uh, and the individual \uh\ values change linearly with the measured UV flux in each epoch.  The model assumes the same super-solar abundances we used in modelling component 1 and in the  photoionization solution for component 3  (see bottom panel of \figref{figPhasePlotComp3AllEpoch}). We further assume that a simple constant covering factor is applicable for all the ionic troughs ($C(v)=0.75$).  Below we give more details for this model and describe its successes and shortcomings.

We modeled component 3a with two Gaussian optical-depth profiles. The velocity centroid of the main Gaussian is at --1880 km s$^{-1}$  and its width is $\sigma=20$ \kms. The high velocity wing was modeled with a Gaussian centered at --1920 km s$^{-1}$ and width of $\sigma=50$ \kms. Component 3b was modeled with a Gaussian centered at --1850 km s$^{-1}$ and width of $\sigma=15$ \kms. As seen in  \figref{figComp3AllEpoch}, the trough shapes in 2002 and 2015 are quite similar when the UV flux  differed by only 34\%. However, large changes are observed between the 2015 and the 2010 epochs, where there is a factor of 2.6 change in the UV flux.  
Therefore,  in  \figref{figComp3model},
we show the best fit results for only the 2010 and 2015 data. 

The main success of this model is the good match for the eight  data sets for the \civ\ and \nv\ troughs (for each ion we fit both doublet components for two epochs: 2010, 2015). This is done with a simple optical depth model where the changes are due to a difference in the ionization parameter of the absorber that equal the flux change between the epochs. 

A reasonable match for the \ovi\ doublet profile in 2015 is obtained using the same parameters that fit the \civ\ and \nv\ troughs (The \ovi\ spectral region was not observed in 2010). 
Quantitatively, the model optical depth is smaller than the data's optical depth by up to 50\% across the trough.  The \Lya\ fit shows similar behavior in the high velocity wing, and an opposite behavior (model deeper than data) in the low velocity wing.  However, it is clear that the AOD assumption for the low velocity wing of \Lya\ is incorrect as the trough does not change in that region between the epochs. The most probable explanation for that is a strong saturation of the \Lya\ trough in that velocity region with a velocity dependent covering factor 
\citep[e.g.,][]{Arav08}.
Another weakness in the model is that, unlike the model for component 1, the best fit model for component 3 invokes changes in \Nh\ for components 3a and 3b by up to a factor of two between the 2010 and 2015 epochs (see \figref{figPhasePlotComp3AllEpoch}).

In summary, most of the troughs' variability in component 3 can be explained by photoionization reaction of two sub components to the quantitative changes in incident ionizing flux, especially the observed back and forth velocity shift of the trough centroid.  However, an AOD model 
with a constant covering factor is inadequate for two reasons. 1) The  \Lya\ trough shows moderate saturation and  a velocity dependent covering factor. 2) In 2002, all 5 troughs from \civ, \nv\ and \Lya\ have a deeper core but shallower high velocity wing (at $v<-1920$ \kms) than the same troughs in the 2010 and 2015 epochs. This behavior is incompatible with pure photoionization changes, as the core and the wing are expected to show depth changes in the same direction. 

\subsubsection{Distance determination from trough variability}

The behavior of component 3's troughs during the 2015 campaign in response to UV flux variation is more complicated than that of component 1.  Different portions of the trough behave in different ways. All 4 troughs from \civ\ and \nv, as well as \Lya\ show trough variation between the June (visit 1) and November (visit 2) 2015 epochs, but only over the narrow velocity range --1910 to -1950 \kms (on the high velocity wing of the trough).  The other portions  of the troughs (full span --1800 to --2070 \kms) do not show variability over this time period.  It is plausible that this behavior is the result of combining moderate saturation across most of the trough with relatively small UV flux changes (a 28\% decrease between visits 1 and 2, see Table 3).  Qualitatively, such a model explains: a) why no variability is observed between the later epochs where the flux changes are less than 16\%;   b) why the entire trough changes between the 2010 and the 2015b epochs (flux change by a factor of 2.6).

The varied response of component 3 to flux variation complicates the extraction of a photoionization time-scale from the data. It is clear that the entire trough shows trough changes correlated with large flux change on a 5 year time scale (between the 2010 and 2015 epochs), while small portions of the trough show such changes on a 6 month timescale (between visits 1 and 2 in 2015).  Similar to our time-dependent photoionization analysis of component 1,
this allows us to derive  log$(\ne)>2.8$ (cm$^{-3}$)  or log$(\ne)>3.6$ (cm$^{-3}$)  using the 5 years and 6 months timescales, respectively. For these estimates, we  used the \nv\ variations, which yields the highest \ne\ 
for the initial conditions given by the photoionization solution for component 3.

In turn, these \ne\ lower limits yield a maximum  distance for component 3 of 150~pc or 60~pc for the 5 years and 6 months time-scales, respectively (where we used the \uh\ value of the global solution for component 3, see Figure 5).

\section{Comparison with the X-ray results}\label{secXray}

As mentioned in section 1, we observed the X-ray manifestation of the outflow with both XMM-Newton and \textit{Chandra}.
Analysis results for the 
The X-ray and UV troughs occupy roughly the same velocity range $-2100<v<-400$ \kms,
see Table 2 here for the UV, Figure 3 in \citet{Peretz18} for the \textit{XMM-Newton} RGS observations, and Table  3 in \citet{Mehdipour18} for the \textit{Chandra} HETG observations.

HST/COS UV data are sensitive to much smaller ionic column densities ($N_{ion}$) than the  XMM/RGS X-ray data. For NGC 7469, the measured UV  CNO $N_{ion}$ are two to three orders of magnitude lower than the secured CNO $N_{ion}$ measurements in the X-ray.  As shown in section 4, the \Lya\ and CNO UV troughs are not heavily saturated.  Therefore, we expect the \Nh\
of the UV phase to be roughly two orders of magnitude smaller than the X-ray phase.

 For the 2015b epoch, we found for component 1 $\log(\Nh)\approx 18.3$ (cm$^{-2}$) and $\log(\uh)\approx -1.0$; and for component 3
$\log(\Nh)\approx 18.8$ (cm$^{-2}$) and $\log(\uh)\approx -1.8$ (see Figures
\ref{figPhasePlotComp1AllEpoch} and \ref{figPhasePlotComp3AllEpoch}).
For the secured CNO $N_{ion}$ measurements in the X-ray, the minimum  $\log(\Nh)$ is between 20--21.3 (cm$^{-2}$), with a spread of $\log(\uh)$ between --0.5 and +0.3. 

These results are in accordance with the expectations outlined in the above paragraph (even when we take into account the super solar metallicity found for the UV material). Accounting for the super-metallicity of the UV photoionization solution, we conclude that in the NGC 7469 outflow, the \Nh\ of the UV material is only about 10\% and 0.5\% of the \Nh\ of the low and high X-ray ionization phases, respectively \citep[see  Figure 6 in][]{Peretz18}.
There is no contradiction between the UV and X-ray results, as the UV data sample material with lower \uh\ values than even the low  ionization X-ray phase.
We note that based on X-ray troughs from ions of Ne, Mg, S and Fe, most of the outflow's \Nh\ 
($\sim10^{22}$~cm$^{-2}$) arises from a much higher ionization phase with  
$\log(\uh)$ between $+0.5$ and $+2$ \citep[see][]{Peretz18,Mehdipour18}.

The UV and X-ray estimates for the distance of the outflow from the central source ($R$) are in agreement and complementary. UV Component 3, which carries most of the \Nh\ of the UV phase, has an upper limit  
of 150 pc or 60 pc (see section 4.3.3). This is consistent with the X-ray warm absorber lower limits for $R$ of 12 pc or 31 pc \citep{Peretz18}.

Assuming the same distance and velocities for the UV and X-ray outflows, we note that the kinetic luminosity of the UV outflowing material will be negligible compared to that of the X-ray component.  This is due to the fact that the kinetic luminosity is proportional to the total column density \citep[see equations 5-7, and accompanied discussion in][]{Borguet12a}. Therefore, since the X-ray phase has a hundred times larger \Nh, it will carry a hundred times larger kinetic luminosity.

\section{Summary}\label{secSummary}

Our multiwavelength campaign on NGC~7469 yielded deep insights about the physical characteristics of this AGN outflow. The UV analysis described in this paper focused on the two major UV absorption components, 1 and 3.
 
Component 1 at $\sim -550$ \kms\, is close to being optically thin in all its troughs, and shows simple photoionization response to changes in the incident ionizing flux. This behavior allowed us to extract a clear physical picture of the absorber: 
\begin{enumerate}
\item The outflowing gas must have about twice proto-solar metallicity. This is similar to our findings for the outflow in Mrk~279 \citep{Arav07}.
\item A simple model based on a fixed total
column-density absorber, reacting to changes in ionizing illumination, matches the observed trough changes in all epochs.  This is similar to our findings for outflow component 1 in NGC~5548 \citep{Arav15}.
\item The simple response to changes of the incident ionizing flux allows us to use time-dependent 
photoionization analysis and obtain a distance of $R=6^{+2.5}_{-1.5}$ pc from the central source.
\end{enumerate}

Component 3 at $\sim -1880$ \kms\ shows a more complicated behavior:  a) Some of its troughs are moderately saturated. b) It shows a complex variability pattern where the velocity centroid shifts by +30 \kms\ in 2010 compared to the 2002 and 2004 epochs, but shifts to its original position in 2015.   A model based on two sub-components reacting to photoionization changes is partially successful in explaining this behavior. Using time-dependent 
photoionization analysis we are able to put an upper limit on its distance $R$ between 60 and 150 pc.

Comparing the physical picture that emerges from the UV analysis to that of the X-ray data we find:  
\begin{enumerate}
\item The total column density of the UV phase is roughly 1\% and 0.1\% of the lower and upper ionization components of the warm absorber, respectively. There is no contradiction here as even the lower  ionization X-ray component has a significantly higher \uh\ value than those inferred for the UV components. 
\item The lower limit on the distance inferred for the X-ray phase is compatible with the upper limits inferred for UV component 3 (that has 3 times higher \Nh\ than component 1).
Together they suggest that the bulk of the outflow in somewhere between 12 and 150 pc from the central source.
\end{enumerate}

Finally, we note that the radial-velocity of all three components is larger than the escape speed at their distance from the AGN.  For the $1.2\times 10^7$ solar masses black-hole in NGC~7469 \citep{Peterson14b}, the escape speed for component 1 at $R=6$ pc, is $\sim130$ \kms\ while its radial
velocity is  $\sim500$ \kms.  Components 2 and 3 are farther away and have higher radial velocities so their
velocities are definitely larger than their escape velocity, even if at their distances the enclosed mass is dominated by the galaxy.  Thus, it is clear that all 3 components are not bound to the gravitational field of the central black-hole.

\vspace{5mm}

\textit{Acknowledgments}. This work was supported by NASA grant NNX16AC07G
through the XMM-Newton Guest Observing Program, and through grants for
HST program number 14054 from the Space Telescope Science Institute, which
is operated by the Association of Universities for Research in Astronomy, Incorporated,
under NASA contract NAS5-26555. The research at the Technion
is supported by the I-CORE program of the Planning and Budgeting Committee
(grant number 1937/12). E.B. received funding from the European Union’s
Horizon 2020 research and innovation program under the Marie Sklodowska-
Curie grant agreement No. 655324. SRON is supported financially by NWO,
The Netherlands Organization for Scientific Research. N.A. is grateful for a
visiting-professor fellowship at the Technion, granted by the Lady Davis Trust.
S.B. and M.C. acknowledge financial support from the Italian Space Agency
under grant ASI-INAF I/037/12/0. B.D.M. acknowledges support from the European
Union’s Horizon 2020 research and innovation program under the Marie
Skłodowska-Curie grant agreement No. 665778 via the Polish National Science
Center grant Polonez UMO-2016/21/P/ST9/04025. L.D.G. acknoweledges support
from the Swiss National Science Foundation. G.P. acknowledges support by
the Bundesministerium für Wirtschaft und Technologie/Deutsches Zentrum fur 
Luft- und Raumfahrt (BMWI/DLR, FKZ 50 OR 1715 and FKZ 50 OR 1812)
and the Max Planck Society. P.O.P. acknowledges support from CNES and from
PNHE of CNRS/INSU. BDM acknowledges support from the European Union’s Horizon 2020 research
and innovation programme under the Marie Skłodowska- Curie grant
agreement No. 798726

\bibliography{astro}

\end{document}